\documentclass{aa}
\pdfoutput=1 
\usepackage{amsmath,amstext}
\usepackage[T1]{fontenc}
\usepackage{txfonts}
\usepackage{graphicx,subcaption}
\usepackage{color}
\usepackage{multirow}
\usepackage{rotating}
\usepackage{pdflscape}
\usepackage{float}

\begin{document}

\title{A  global view of the inner accretion and ejection flow around super massive black holes}

   \subtitle{Radiation-driven accretion disk winds in a physical context}

   \author{Margherita Giustini
          \inst{1,2}
          \and
          Daniel Proga\inst{3}           }

   \institute{Centro de Astrobiolog\'ia (CSIC-INTA), Dep. de Astrof\'isica; ESAC campus, Villanueva de la Ca\~nada, E-28692 Madrid, Spain\\
   \and
   SRON Netherlands Institute for Space Research, Sorbonnelaan 2, 3584 CA Utrecht, The Netherlands\\
                     \and
                Department of Physics and Astronomy, University of Nevada Las Vegas, NV 89154 Las Vegas, USA          }

   \date{Received 10 July 2018 / Accepted 11 April 2019}

\abstract
   {Understanding the physics and geometry of accretion and ejection around super massive black holes (SMBHs) is important to understand the evolution of active galactic nuclei (AGN) and therefore of the large scale structures of the Universe.}
    {We aim at providing a simple, coherent, and global view of the sub-parsec accretion and ejection flow in AGN with varying Eddington ratio, $\dot{m}$, and black hole mass, $M_{BH}$.}
   {We made use of theoretical insights, results of numerical simulations, as well as UV and X-ray observations to review the inner regions of AGN by including different accretion and ejection modes, with special emphasis on the role of radiation in driving powerful accretion disk winds from the inner regions around the central SMBH.}
   { We propose five $\dot{m}$ regimes where the physics of the inner accretion and ejection flow around SMBHs is expected to change, and that correspond observationally to quiescent and inactive galaxies; low luminosity AGN (LLAGN); Seyferts and mini-broad absorption line quasars (mini-BAL QSOs); narrow line Seyfert 1 galaxies (NLS1s) and broad absorption line quasars (BAL QSOs); and super-Eddington sources.
   We include in this scenario radiation-driven disk winds, which are strong in the high $\dot{m}$, large $M_{BH}$ regime, and possibly present but likely weak in the moderate $\dot{m}$, small $M_{BH}$ regime. }
   {A great diversity of the accretion/ejection flows in AGN can be explained to a good degree by varying just two fundamental properties: the Eddington ratio $\dot{m}$ and the black hole mass $M_{BH}$, and by the inclusion of accretion disk winds that can naturally be launched by the radiation emitted from luminous accretion disks.
   }

\keywords{black hole physics --- galaxies: active --- galaxies: nuclei --- quasars: general --- quasars: super massive black holes}

\titlerunning{A global view of accretion and ejection around SMBHs}
\authorrunning{Giustini \& Proga}

\maketitle
\section{Introduction}\label{SECTION:INTRO}
Active galactic nuclei (AGN) are powered by mass accretion onto super massive black holes (SMBHs) with masses $M_{BH}\sim 10^6-10^{10} M_{\odot}$,  and are crucial to our understanding of the formation and evolution of the cosmic structures \citep[e.g.,][]{1998A&A...331L...1S,2006ApJS..163....1H, 2012ARA&A..50..455F}.
The luminosity produced by mass accretion can be written as $L=\eta \dot{M}c^2$, where $\eta$ is the accretion efficiency, $\dot{M}$ the mass accretion rate, and $c$ the speed of
light. The mass accretion rate can be rescaled for the black hole (BH) mass through the
dimensionless Eddington ratio parameter: $\dot{m}\equiv \dot{M}/\dot{M}_{Edd}\equiv
L/L_{Edd}$, where $L_{Edd}$ is the Eddington luminosity.
Historically, local AGN are referred as Seyfert galaxies \citep{1943ApJ....97...28S} and more distant luminous AGN as quasi-stellar objects, quasars, or QSOs \citep{1963Natur.197.1040S}. We will refer to both Seyferts and QSOs as (luminous) AGN, differentiating among them only on the basis of different physical properties such as BH mass and luminosity, which are higher in QSOs than in Seyferts.
The observed AGN bolometric luminosity spans a wide range: from $L\sim 10^{40-42} $ erg s$^{-1}$ in low-luminosity AGN \citep[LLAGN,][]{1999ApJ...516..672H}, to $L\sim 10^{43-46} $ erg s$^{-1}$ in Seyfert galaxies and low-luminosity QSOs  \citep[e.g.,][]{2012MNRAS.425..623L}, to $L\sim 10^{44-48} $ erg s$^{-1}$ in high-luminosity QSOs \citep[e.g.,][]{2011ApJS..194...45S}; therefore implying a wide range of either $\dot{m}$, or $\eta$, or both.

The spectral energy distribution (SED) of AGN covers the full electromagnetic
spectrum \citep{1994ApJS...95....1E,2011ApJS..196....2S}, and in particular the optical, UV,
and X-ray portions of the SED originate from the regions closest to the SMBH, on sub-parsec scales \citep[e.g.,][]{1997iagn.book.....P,1993ARA&A..31..717M}.
Observations from the optical to the X-rays show that the AGN SED varies with $\dot{m}$, giving relatively more X-ray photons compared to optical and UV  ones with increasing $\dot{m}$ in LLAGN  \citep{2011ApJ...739...64X}, and relatively less X-ray photons compared to lower energy ones in luminous AGN with increasing $\dot{m}$ \citep[e.g.,][]{2009MNRAS.392.1124V,2012MNRAS.425..907J}.
The distinctive optical and UV observational features of luminous AGN, the big blue bump, and the broad line region (BLR) are absent or very weak in LLAGN; LLAGN also lack the typical large variability on short timescales displayed in the X-ray band by luminous AGN \citep{2008ARA&A..46..475H}. LLAGN and luminous AGN should be powered by physically different inner accretion flows: a hot, geometrically thick, radiatively inefficient flow in the former case \citep{2014ARA&A..52..529Y}, and a cold, geometrically thick, radiatively efficient flow in the latter case \citep{1973A&A....24..337S}.
The mutual interaction of the AGN with its environment, or feedback, is expected to be mainly kinetic in the case of LLAGN, via their highly collimated, relativistic polar jets \citep[e.g.,][]{2012ARA&A..50..455F}, while it is expected to be mainly in the form of radiation (radiative) in the case of luminous AGN \citep[e.g.,][]{2014ARA&A..52..589H}.
However, it has become clear in the past few decades that also in the case of luminous AGN a substantial amount of energy can be released as kinetic luminosity, through massive, sub-relativistic, wide-angle winds \citep{1998A&A...331L...1S}.

Wide-angle, sub-relativistic winds originating from the inner regions around SMBHs are commonly observed as blueshifted absorption lines in the UV and X-ray spectra of luminous AGN \citep{1991ApJ...373...23W,2009ApJ...692..758G,2010A&A...521A..57T}.
At least half of local AGN show the presence of low-velocity narrow\footnote{Historically, narrow absorption lines (NALs) are defined to have a Full Width at Half Maximum (FWHM)
 $< 500$ km s$^{-1}$), while broad absorption lines (BALs) have a FWHM $> 2000$ km s$^{-1}$; the intermediate cases are called mini-broad absorption lines \citep[mini-BALs; see e.g.,][]{1981ARA&A..19...41W,2004ASPC..311..203H,2009ApJ...696..924G}. } absorption lines with blueshift of a few $100-1000$ km s$^{-1}$, both in the UV and in the X-ray band\footnote{In the soft X-ray band, the plethora of narrow absorption lines are called collectively ``warm absorber'' \citep[e.g.,][]{1995MNRAS.273.1167R}.} \citep{2003ARA&A..41..117C,2014MNRAS.441.2613L}. Their ionization state ranges from, for example, Mg~\textsc{ii} to C~\textsc{iv} and O~\textsc{vi} for the UV NALs, to, for example, O~\textsc{viii} for the X-ray NALs \citep[e.g.,][]{2010SSRv..157..265C}.
Similar ionization states of those low-velocity UV NALs, but much higher terminal velocities up to $\sim 0.2c$ and much broader features, are observed in about 20-30\% of optically selected AGN in the form of UV BALs and mini-BALs \citep[e.g.,][]{1991ApJ...373...23W,2006ApJS..165....1T,2008MNRAS.386.1426K,2009ApJ...692..758G}.
Even higher ionization states and velocities, for example, Fe~\textsc{xxv}, Fe~\textsc{xxvi} blueshifted by $0.1-0.4c$, are inferred for the X-ray absorbing ultrafast outflows (UFOs), also observed in a large fraction of local AGN \citep[$\sim 20-40\%$,][]{2006AN....327.1012C,2010A&A...521A..57T,2013MNRAS.430...60G}.
The few detailed X-ray spectroscopic analyses that have been performed on high redshift sources, revealed numerous cases of highly ionized outflowing absorbers \citep[][]{2002ApJ...579..169C,2003ApJ...595...85C,2012A&A...544A...2L,2014ApJ...783...57C,2015A&A...583A.141V,2016ApJ...824...53C,2018A&A...610L..13D}.
These findings suggest that high-velocity winds are widespread in terms of both geometrical covering factor and rate of occurrence also among high redshift, luminous AGN.

Theoretically, mass outflows can be launched either by thermal pressure, magnetic forces, or radiation. Independent of the mechanism of wind launching, in order to be successfully launched, a wind has to overcome the SMBH gravitational pull. The wind terminal velocity is proportional to the escape velocity $\upsilon_{esc}(R)=\sqrt{2GM_{BH}/R}$, therefore the closer the central SMBH is to the wind launching region, the faster the wind terminal velocity will be \citep[e.g.,][]{2007ASPC..373..267P}.
Given the masses and temperatures involved, thermal winds in AGN can be launched only at large radii, $R\gtrsim 10^{5} R_g$ ($R_g\equiv GM_{BH}/c^2$ is the gravitational radius), from the central SMBH, and can reach terminal velocities of the order of a few $100-1000$ km s$^{-1}$ \citep{2001ApJ...561..684K,2008ApJ...675L...5D}.
Magnetocentrifugally-driven winds with similar terminal velocities can be launched from similarly large scales \citep{2012ApJ...761..130Y}.
Radiation pressure on dust grains can launch winds on radial scales larger than the dust sublimation radius \citep{2006MNRAS.373L..16F,2008MNRAS.385L..43F}, and shape the environment of AGN by determining their obscuration at such scales \citep{2017Natur.549..488R}.
To launch more powerful winds with terminal velocities $\gg 2000$ km s$^{-1}$ and up to significant fractions of the speed of light, radiation pressure or magnetic forces acting on smaller (sub-parsec) scales must be at work instead \citep{2000ApJ...543..686P,2015ApJ...805...17F}.

We present a unified picture of the relevant theory and observations pertaining to the inner, sub-parsec scale accretion and ejection flow around SMBHs with varying $\dot{m}$ and, secondarily, $M_{BH}$. We emphasize the role of radiation in launching winds from accretion disk scales in luminous AGN.
The Eddington ratio $\dot{m}$ is likely the main physical driver of physical differences in the accretion flow that lead to different observed phenomenology of AGN \citep[e.g., the Eigenvector 1:][]{2000ApJ...536L...5S,2001ApJ...558..553M,2002ApJ...565...78B,2014Natur.513..210S}.
Generally speaking, $\dot{m}$ is proportional to the density of the matter involved in the accretion flow around the SMBH.
The density of the accretion flow regulates the relative importance of matter heating and cooling, and through this the consequent geometrical and physical structure of the mass accretion and ejection flow itself, along with the correspondent flow of photons, or SED. The AGN SED, which can also strongly depend on $M_{BH}$, can in turn affect the inner mass accretion and ejection flow structure through photoionization, absorption, and scattering.
In particular, UV radiation can induce further mechanical feedback by driving powerful line-driven (LD) winds from accretion disk scales \citep[][hereafter PK04]{1995ApJ...451..498M,2000ApJ...543..686P,2004ApJ...616..688P}.

We present our global scenario for the inner accretion and ejection flow around SMBHs in Sect.~\ref{MAIN}, the comparison with observations in Sect. \ref{OBS}, the discussion in Sect.~\ref{DISC}, and we summarize and conclude in Sect.~\ref{CONC}.

 \section{The inner accretion and ejection flow in active galactic nuclei, re-viewed\label{MAIN}}
\begin{figure*}[h!]
\centering
\includegraphics[width=12cm]{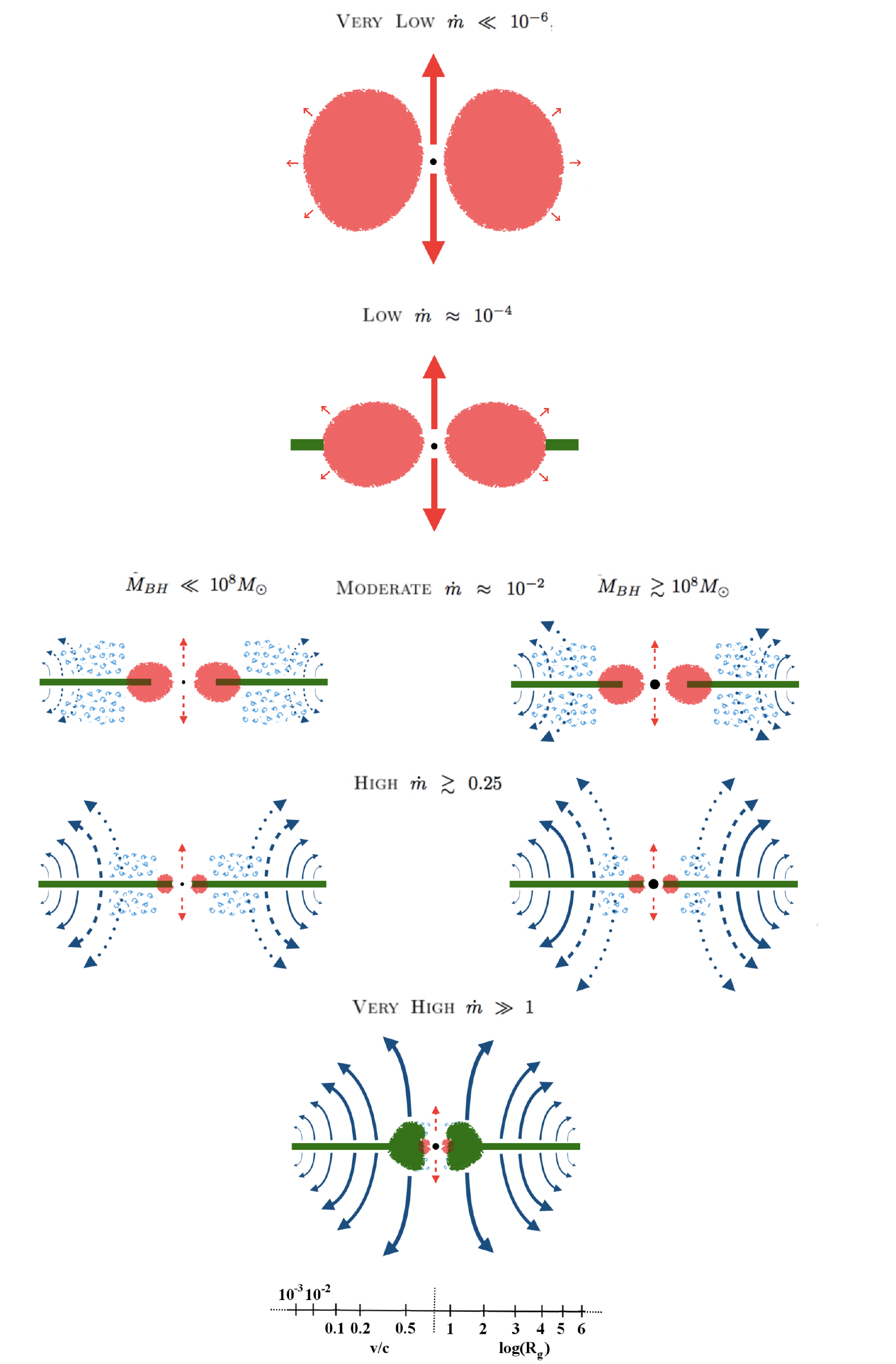}
   \caption{Logarithmic-scale side view of the inner parsec of the accretion and ejection flow of AGN for five regimes of $\dot{m}$ that increase from top to bottom.
We plot in light red the hot, optically thin accretion flow; in green the cold, optically thick accretion flow; in red we plot the magnetically driven ejection flow streamlines; in  blue the radiation-driven accretion disk wind and failed wind streamlines.
In the optically thick, geometrically thin disk dominated cases, two extremes of very small (left) and very large (right) $M_{BH}$ are presented.
The length and thickness of the arrows reflect the strength of the wind in terms of terminal velocity. From the closest to the farthest arrow to the central SMBH, velocities decrease from $\sim c$ of the radio jet, to the  $\sim 0.4c$ of the fastest UFOs launched at $r\sim 10 r_g$, to the $\sim 0.1-0.2c$ typical of BALs and mini-BALs launched at $r\sim 10^2-10^3 r_g$, to $\sim 0.01-0.1c$ of the BALs and mini-BALs launched at $r\sim 10^3-10^4 r_g$, out to $\sim 0.001c$ of low-velocity NALs launched at $r>10^4 r_g$.
Solid lines represent streamlines of persistent winds, while dashed and dotted lines represent streamlines of transient, sporadic ejection flows; the blue wiggly little clouds represent flow streamlines of the completely failed wind.
The actual duty cycle of the wind is still uncertain, and its determination is  the subject of current observational and theoretical efforts.}
         \label{FIGURA}%
    \end{figure*}
    \begin{figure*}[h!]
    \centering
\includegraphics[width=11cm]{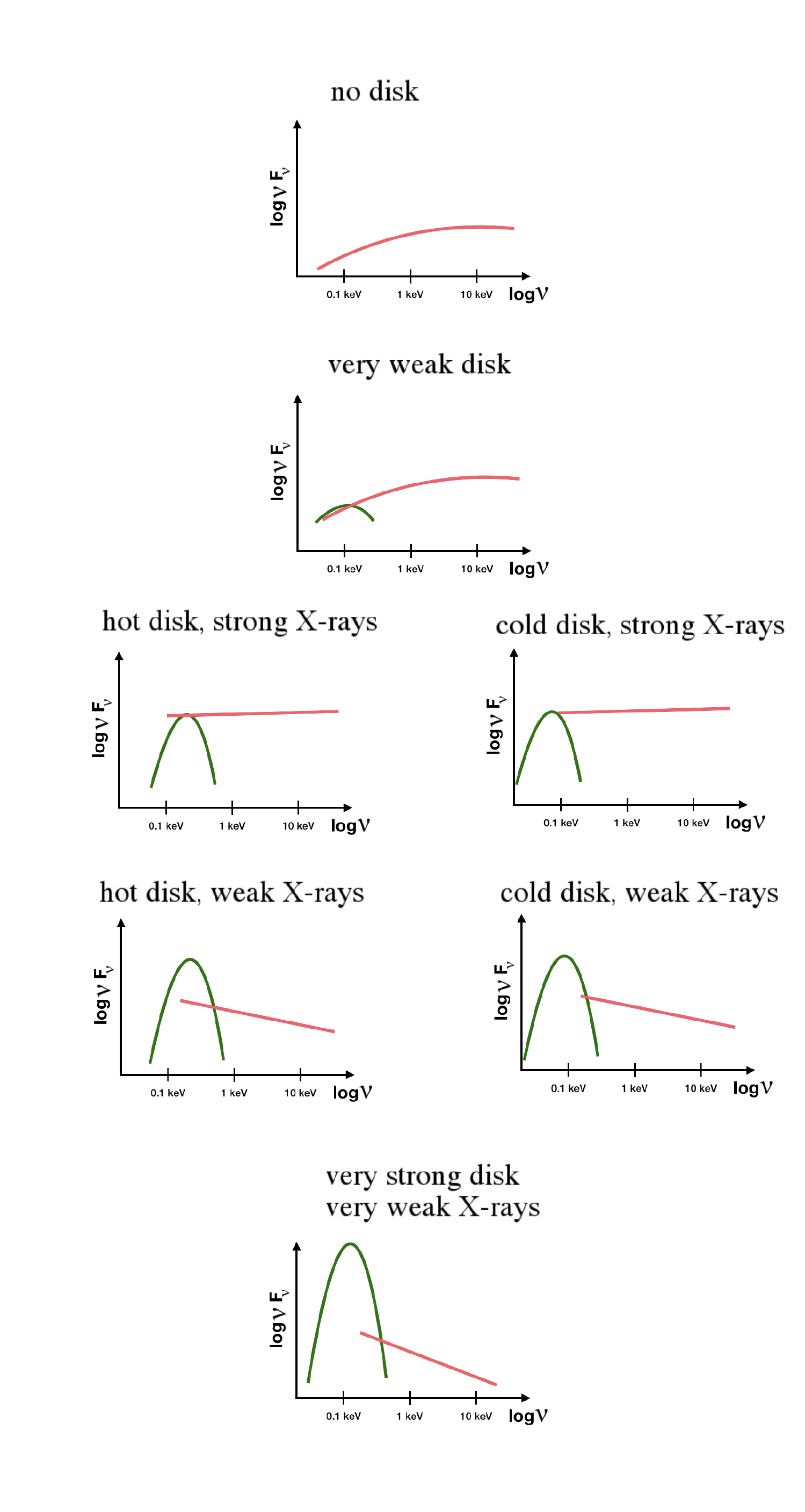}
   \caption{Sketch of the optical to X-ray AGN SED for the seven different $\dot{m},M_{BH}$ cases presented in Fig.~\ref{FIGURA}. The non-thermal emission due to the hot optically thin matter is plotted in red, the thermal emission due to the cold optically thick matter is plotted in green.
   }    \label{FIGURAsed} %
    \end{figure*}

We consider different values of the Eddington ratio, from very low $\dot{m}\ll 1 $ up to very high $\dot{m}\gg 1 $, and identify three intermediate regimes (low, moderate, and high $\dot{m}$), for a total of five  $\dot{m}$ regimes, where the physics of accretion/ejection around SMBHs is expected to significantly change.
This is equivalent to following the evolution of a hot, optically thin accreting plasma that increases its density and cools down by emitting radiation, becoming more and more optically thick with increasing $\dot{m}$.
The accretion radiative efficiency $\eta$ depends on the radiative cooling rate and this depends, among other parameters, on the density of the accretion flow and on the BH spin.
The dependence of the radiative efficiency on the density of the accretion flow can be translated into a dependence on $\dot{m}$: for $\dot{m}\lesssim 1$, $\eta$ is directly proportional to $\dot{m}$ \citep{2014ARA&A..52..529Y}, while at $\dot{m}\gg 1$ a significantly larger part of accretion energy gets advected into the SMBH before being radiated away \citep[see, e.g.,][for a recent review]{2018arXiv180706243M}.

Theoretically, at very low $\dot{m}\ll 10^{-6} $, the accretion flow is very tenuous, optically thin, and the cooling is negligible. Therefore, the flow is hot and radiatively inefficient ($\eta\sim 0$), and geometrically thick \citep{1995ApJ...452..710N,2014ARA&A..52..529Y,2015MNRAS.454.1848R}.
At low $\dot{m}\approx 10^{-4}$, the radiative efficiency is no longer negligible  ($\eta\sim 10^{-3}$) and the outer parts of the accretion flow start to cool down \citep{2017MNRAS.466..705S,2017ApJ...844L..24R}.
Above a moderate $\dot{m} \gtrsim 0.01$, the cooling is so rapid that the accretion flow is radiatively efficient ($\eta\sim 0.1$), optically thick, and geometrically resembles a thin disk \citep{1973A&A....24..337S}.
While at low and very low $\dot{m}$, the ejection flow is dominated by collimated relativistic polar jets \citep{2014ARA&A..52..529Y}, at moderate $\dot{m}$,  the jets are much weaker, and more equatorial winds originating from the accretion disk can develop \citep{1995ApJ...451..498M}. It is only at high $\dot{m}\gtrsim 0.25$, however, that these equatorial, sub-relativistic winds originating on accretion disk scales can dominate the ejection flow (PK04).
Eventually, when the Eddington ratio goes well above the Eddington limit at very high $\dot{m}\gg 1$, the inner accretion flow becomes radiation pressure dominated, changing its geometrical configuration from a  thin disk to a thicker one; the covering factor of the sub-relativistic winds will be maximum, and also polar outflows will be present \citep[e.g.,][]{2016MNRAS.456.3929S}.

We assume that the higher $\dot{m}$, the larger the radial extent of the outer cold and optically thick accretion flow. In other words, the cold flow moves further in toward the central SMBH, whereas the inner accretion flow that is hot and optically thin has its size and temperature decreasing with increasing $\dot{m}$.
This is similar to the models of \citet{2004A&A...414..895F} for LLAGN, and of \citet{1997ApJ...489..865E} and \citet{2007A&ARv..15....1D} for stellar mass black hole binaries, and to the model by \citet{2000A&A...360.1170R} for accreting BHs in general.
As for the outflow, we assume that with increasing $\dot{m}$, it becomes less and less dominated by the polar relativistic jets and more and more by the sub-relativistic accretion disk winds, and that these winds increase their geometrical covering factor with increasing $\dot{m}$.

\begin{table*}
\caption{Summary of the main properties of the five $\dot{m}$ regimes sketched in Fig.~\ref{FIGURA} and described in Sects. ~\ref{SEC:VL} to ~\ref{SEC:VH}. }
\label{TABLE}
\centering
\begin{tabular}{c c c c c}     
\hline\hline
 $\dot{m}$ range & Accretion/ejection flow &      Feedback &      Examples \\
(1) & (2) & (3) & (4)\\
\hline
very low $\dot{m}\approx 10^{-8}$ & non-radiative hot accretion flow & \multirow{2}{*}{$L_{kin}$} & Quiescent/inactive, \\
($ \ll 10^{-6}$)             &  relativistic polar jet   & & Sgr A*\\
 & & & \\
low $\dot{m}\approx 10^{-4}$  & outer cold disk at $\sim $1000s $R_g$, inner hot flow &  \multirow{2}{*}{$L_{kin} \gg L_{rad}$} & LLAGN \\
 ($  10^{-6} \lesssim \dot{m} \lesssim10^{-3}$) & relativistic polar jet & &  M81*, M87\\
 & & & \\
 moderate $\dot{m}\approx 10^{-2}$  &  outer cold disk at $\sim$10s $R_g$, extended hot corona & \multirow{2}{*}{$L_{kin}\ll L_{rad}$} & Seyfert/mini-BAL QSO \\
 ($10^{-3} \lesssim \dot{m} \lesssim10^{-1}$) & weak/moderate LD wind depending on small/large $M_{BH}$ &  & NGC 5548/PG 1126-041 \\
 & & &\\
 high $\dot{m}\gtrsim 0.25$  & cold accretion disk down to ISCO, compact hot corona & \multirow{2}{*}{$L_{kin} <  L_{rad}$} & NLS1/BAL QSO \\
 ($0.1\lesssim \dot{m} \lesssim 1$) & moderate/strong LD wind depending on small/large $M_{BH}$ & &  I Zw 1/PDS 456 \\
& & &\\
 very high $\dot{m}\gg 1$ & outer thin disk, inner slim disk, very compact hot corona & \multirow{2}{*}{$L_{kin}\lesssim L_{rad}$} &Super-Eddington  \\
 ($ 1\lesssim \dot{m} \lesssim 100$) & strong outflows, both polar and equatorial &  &  RX J0439.6-531\\
    \hline
\end{tabular}
\tablefoot{(1) Nomenclature for the Eddington ratio ranges used in this work, with an indicative order of magnitude, and an indicative range of values in parentheses. (2) Accretion and ejection flow main physical characteristics. (3) Type of energy feedback between the AGN and the environment: kin = kinetic, rad = radiative. (4) Classes of objects or individual examples of well-studied local AGN.}
\end{table*}

The main properties of the structure of the inner accretion/ejection flow for the five different $\dot{m}$ regimes are summarized in Table~\ref{TABLE} along with examples of known AGN for each regime.
In Fig.~\ref{FIGURA}, we present a sketch of the side view of the  inner parsec AGN structure for the five different $\dot{m}$ regimes.
Figure~\ref{FIGURAsed} presents a sketch of the intrinsic SED corresponding to each $\dot{m}/M_{BH}$ regime. To  illustrate the effects of the different accretion disk temperature on the structure of the radiation driven winds for different BH masses, for the luminous AGN in the moderate and high $\dot{m}$ regime we present the two cases: $M_{BH}\ll 10^8 M_{\odot}$ and $M_{BH}\gtrsim 10^8 M_{\odot}$.

In Sect. \ref{SEC:BHMASS} we discuss the effects of varying black hole mass in luminous AGN, while in the following subsections, we describe the main physical and observational properties of the five different $\dot{m}$ regimes:
very low $\dot{m}\ll 10^{-6} $ in Sect.~\ref{SEC:VL}; low $10^{-6}\lesssim\dot{m}\lesssim 10^{-3}$ in Sect.~\ref{SEC:L}; moderate $\dot{m}\gtrsim 10^{-2}$ in Sect.~\ref{SEC:M}; high $\dot{m}\gtrsim 0.25$ in Sect.~\ref{SEC:H}; and very high $\dot{m}\gg 1$ in Sect.~\ref{SEC:VH}.

\subsection{Effect of black hole mass \label{SEC:BHMASS}}
    \begin{figure}[h!]
\centering
\includegraphics[width=8.5cm]{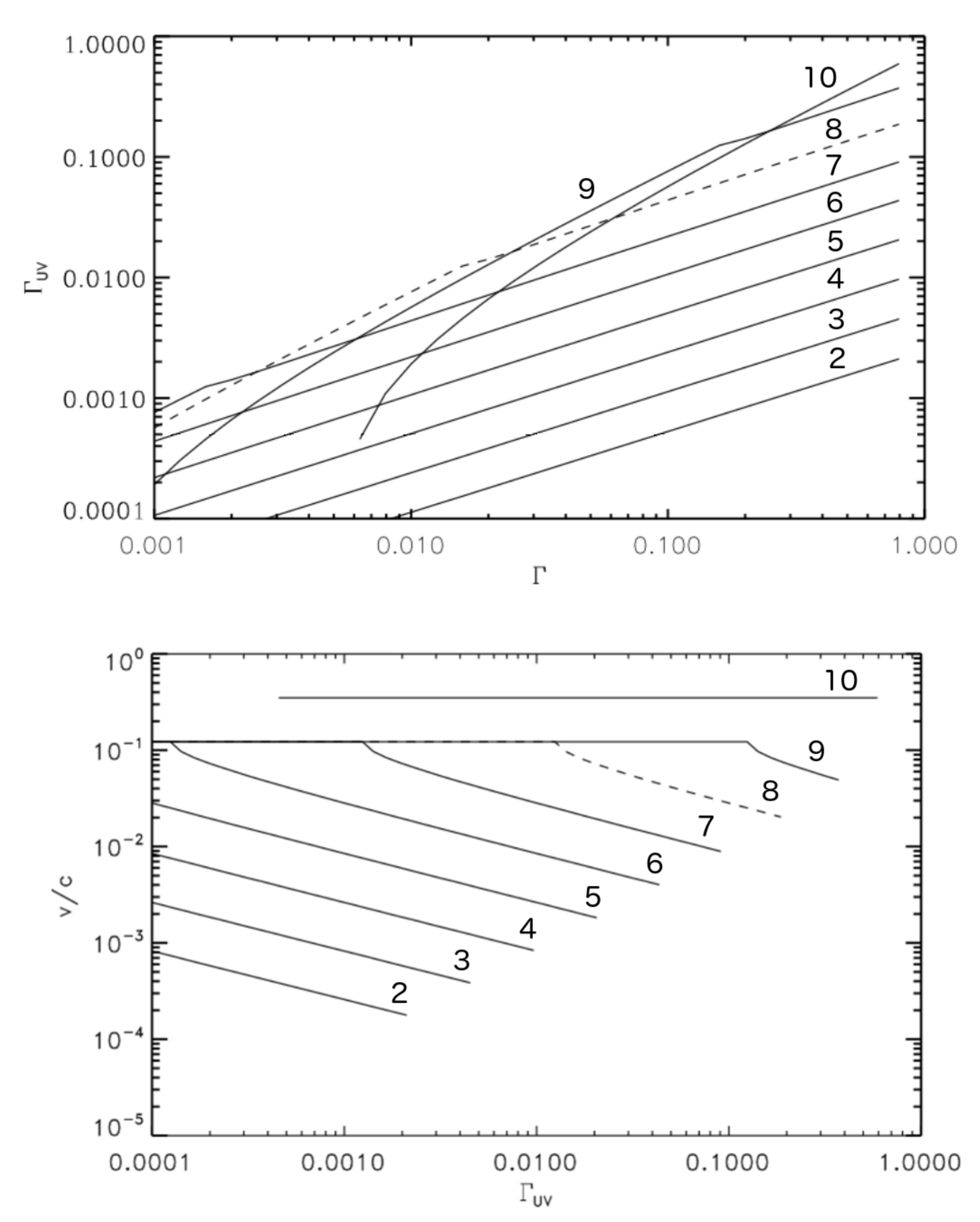}
   \caption{Top panel: UV Eddington luminosity  $L_{UV}/L_{Edd}\equiv \Gamma_{UV}$ as a function of bolometric Eddington luminosity $L/L_{Edd}\equiv \Gamma$ for a \citet{1973A&A....24..337S} accretion disk around nine different black hole masses, from $10^2 M_{\odot}$ to $10^{10} M_{\odot}$; the case $M_{BH}=10^8 M_{\odot}$ is plotted with a dashed line.
   Bottom panel: Circular velocity at the radius where the disk temperature $T=50,000$ K or $T = T_{max}$ if $T_{max} < 50,000$ K, as a function of $\Gamma_{UV}$, for the same nine values of $M_{BH}$ as in the top panel. }      \label{FIGURAgamma}%
    \end{figure}
At the lowest $\dot{m}$, matter is fully ionized, and the accretion/ejection flow properties are independent of the BH mass, as the plasma physics governing the corresponding flows is scale invariant.
Scaling of the accretion and ejection flow physical properties across the mass range is broken once the density increases enough to allow for cooling to become efficient, and the atomic absorption opacities of the accreted/ejected matter to become important.
In particular, in the \citet{1973A&A....24..337S} solution for $\dot{m}\gtrsim 0.01$, where the accretion flow is a geometrically thin, optically thick accretion disk, the temperature scales as $T^4\propto \left(\dot{m}/M_{BH}^2\right)\left(R_{in}/R_g\right)^{-3}$.
For a large $M_{BH}\gtrsim 10^8 M_{\odot}$, the peak disk temperature will be around the optical/UV; while for small $M_{BH}\ll 10^8 M_{\odot}$, it will move toward the far UV/soft X-ray regime.

The relative UV/X-ray photon flux and matter opacities of the inner accretion flow are critical to the launching and acceleration of line-driven accretion disk winds (LD disk winds). LD acts as a booster of the radiation pressure when most of the gas opacity is in spectral lines, and it is therefore strongly dependent on the matter ionisation state \citep{1975ApJ...195..157C,1995ApJ...451..498M,2000ApJ...543..686P, 2018arXiv181201773D}.
A UV flux that is too low will not be able to exert enough pressure to launch a LD wind, and a X-ray flux this is too large will ionize the matter above a level where LD is no longer effective  (``overionizing'' it).
Therefore at a given $\dot{m}$, a large $M_{BH}$ will favor the development of a powerful (dense, fast, persistent) LD disk wind over a larger range of radial disk scales, while a small $M_{BH}$ will have a disk  where the circumnuclear gas over a large range of radial scales will get overionized and will therefore be a failed LD disk wind \citep[][PK04]{2002ApJ...565..455P,2005ApJ...630L...9P}.

Following \citet{2002ASPC..255..309P}, we plot in the top panel of Fig.~\ref{FIGURAgamma} the UV Eddington ratio $L_{UV}/L_{Edd}$ as a function of the bolometric Eddington ratio $L/L_{Edd}$, for nine different black hole masses, from $10^2 M_{\odot}$ to $10^{10} M_{\odot}$.
To calculate $L_{UV}$, a pure \citet{1973A&A....24..337S} accretion disk was assumed, and its radial temperature profile was calculated where the disk temperature $12,000$ K $\leq T \leq  50,000$ K (or  $12,000$ K $\leq T \leq T_{max}$ if $T_{max} < 50,000$ K).
The UV photons are the main ones capable of driving a LD wind, while higher energy photons coming from hotter regions of the accretion disk are the ones capable of making the wind fail, by overionizing the UV-absorbing atoms.
Focusing on the $M_{BH}=10^8 M_{\odot}$ case, which is marked by the dashed line in Fig.~\ref{FIGURAgamma}, one can see the effect of increasing disk temperature with increasing $\dot{m}$: at 1\% of the Eddington limit almost all the luminosity is emitted in UV ($\Gamma_{UV} \lesssim \Gamma$); at $\Gamma\sim 0.1$, $L_{UV}$  is about a half of the total disk luminosity; and at $\Gamma\sim 0.5$, $L_{UV}$ is about one fifth of the total luminosity. For the case of much smaller $M_{BH}=10^6 M_{\odot}$, the temperatures involved are much hotter than the $M_{BH}=10^8 M_{\odot}$ case, and the corresponding  $L_{UV}$ is less relevant to the bolometric budget, being $\sim 1/5, 1/10,$ and $1/15$ of the total disk luminosity when $\Gamma\sim 0.01, 0.1,$ and $0.5$.

In the case of SMBHs, the temperatures will allow the development of LD accretion disk winds (PK04).
In the bottom panel of  Fig.~\ref{FIGURAgamma}, we plot the circular velocity $\upsilon(R)=\sqrt{GM_{BH}/R}$, where $R$ is the radius where $T=50,000$ K or $T = T_{max}$ if $T_{max} < 50,000$ K, as a function of $\Gamma_{UV}$, for the same nine values of $M_{BH}$ as above. The LD disk wind terminal velocity is $\upsilon_{out}\sim 4\upsilon$.
In this figure one can see how large terminal velocities  $\upsilon_{out}> 0.1c$ are reachable in principle through LD at large BH masses.
These values are computed taking into account only UV radiation, while the actual presence of ionizing X-ray photons will move down the curves for the terminal velocities at all masses.
The important point is that given the SED of AGN, when $\dot{m}$ and $M_{BH}$ are large enough, the launch and acceleration of a LD accretion disk wind is expected \citep[PK04;][]{2010A&A...516A..89R}.

\subsection{Very low $\dot{m} \ll 10^{-6}$\label{SEC:VL}: Quiescent active galactic nuclei}
In this $\dot{m}$ regime, radiation-driven winds cannot be launched.
From a theoretical perspective, at very low $\dot{m}$ (much less than $10^{-6}$), the density is so low that electron heating is negligible, most of the accretion energy goes into the ions that do not emit significantly, and causes them to be advected into the SMBH \citep{1995ApJ...452..710N}.
At such low densities the radiative cooling, and thus the accretion radiative efficiency $\eta$, are negligible, and
the inner accretion flow is hot, optically thin, and geometrically thick. High velocity bipolar outflows, that is, jets, are expected \citep{1994ApJ...428L..13N,1999MNRAS.303L...1B}.
Numerical simulations show that the flow can, however, deviate easily from being advection dominated,
and start to emit with some non-null radiative efficiency. In particular, when heating is allowed
to evolve independently in ions and electrons, most of the heat
goes into the electrons in the more polar regions of the flow that are dominated by magnetic pressure,
with respect to the more equatorial regions that are gas-pressure dominated and where most
of the heat goes into the ions.
Albeit non-negligible, the radiative efficiency is very low ($\eta\ll 0.1\%$), and the kinetic luminosity associated to the jet completely dominates the AGN energetic output \citep[e.g.,][]{2015MNRAS.454.1848R}.
The continuum SED is dominated by bremsstrahlung and Compton emission from the hot optically thin flow, where the energetic contribution of the latter increases over the former with increasing $\dot{m}$ \citep{1997ApJ...477..585M}, and by synchrotron emission from the polar jet \citep[e.g.,][]{2014A&A...570A...7M}.
Thermal emission features are absent, and intrinsic obscuring structures such as a torus or a BLR are not expected to exist \citep{2000ApJ...530L..65N,2006ApJ...648L.101E,2009ApJ...701L..91E}.
Low-velocity ($\upsilon_{out}\sim 100-1000 $ km s$^{-1}$) magnetically or thermally driven winds can be launched in the outer regions of the hot accretion flow, on the order of tens of thousands $R_g$ \citep[e.g.,][]{2012ApJ...761..130Y}. These winds would appear as slightly blueshifted, weak (column density $N_H\lesssim 10^{20}$ cm$^{-2}$) X-ray NALs.
Systems in this regime are quiescent or inactive, such as our own Galactic center Sgr A*.

\subsection{Low $\dot{m} \sim 10^{-4}$\label{SEC:L}: Low-luminosity active galactic nuclei}
As the density of the inner accretion flow increases, radiative cooling starts to be important in the more equatorial regions of the flow, and the hot optically thin flow collapses in its outer zones into a colder, geometrically thinner zone: this will happen around $\dot{m}\sim 10^{-4}$ \citep{2017MNRAS.466..705S,2017ApJ...844L..24R} and at radii on the order of $10^{3}-10^4 R_g$ \citep{2014ARA&A..52..529Y}.
At this low $\dot{m}$, the electrons can be already efficiently heated up in the polar regions by magnetic fields, and radiation coming from these hot regions can be produced with an efficiency as high as $\eta\sim 0.005$ \citep{2007ApJ...667..714S} or even higher, depending on the electron temperature, but always lower than $\eta\sim 0.05$ \citep{2017MNRAS.468.1398S}.
At these low Eddington ratios, the torus, the BLR, and their associated reprocessing features are still negligible or very weak, and as in the previous $\dot{m}$ regime most of the SED will be due to the Compton, bremsstrahlung, and synchrotron emission from the inner hot flow and the jet.

Systems in the low $\dot{m}$ regime are LLAGN, which lack strong thermal emission in the optical/UV \citep{1999ApJ...516..672H} and where the disk, if present, is truncated at large radii.
The contribution of the thermal disk to the SED will be therefore very small, albeit non-null.
As in the previous regime, the ejection flow is magnetically dominated, with high-velocity collimated polar radio jets that through kinetic luminosity exert most of the feedback, with the addition of low-velocity outer winds \citep{2012ApJ...761..130Y, 2014ARA&A..52..529Y}.
The outer winds can imprint weak features in the UV and X-ray spectra as low-velocity, low column density ($\upsilon_{out}\sim 100-1000 $ km s$^{-1}$, $N_H\sim 10^{20-21}$ cm$^{-2}$) NALs.
Increasing $\dot{m} $ will decrease the optically thick equatorial region truncation radius and increase its thermal contribution in the optical/UV SED, going toward the next $\dot{m}$ regime.

\subsection{Moderate $\dot{m}\sim 10^{-2}$\label{SEC:M}: Luminous active galactic nuclei}
In this $\dot{m}$ regime, radiation pressure is able to launch winds from accretion disk scales.
Most of the accretion flow above $\dot{m}\sim 10^{-2}$ is optically thick and radiatively efficient \citep[$\eta\sim 0.1$,][]{2011ApJ...728...98D}, as the accreted matter is able to radiatively cool down and stay geometrically thin: theoretically, the accretion flow transitions into an accretion disk that emits mainly in the optical and UV, with a peak temperature inversely proportional to the BH mass \citep{1973A&A....24..337S}.
There are no advanced and extensive numerical simulations yet for the transition of AGN from the previous into this Eddington ratio regime, where atomic physics effects severely complicate the computations. Therefore the exact geometrical configuration of the inner accretion/ejection flow is largely uncertain \citep{2017MNRAS.468.1398S}.

Here we assume that the accretion disk does not extend always down to the innermost stable circular orbit (ISCO) but that instead the optically thick and optically thin  matter distribution smoothly follows the configuration in the previous low $\dot{m}$ regime, with the optically thick disk extending more and more toward the central SMBH with increasing $\dot{m}$, and toward the next high $\dot{m}$ one, where the disk finally reaches ISCO.
The most central portion of the accretion flow consists of the hot, compact, optically thin plasma \citep[in this regime called the X-ray corona, e.g.,][]{1991ApJ...380L..51H} that Compton upscatters a fraction of the thermal seed disk photons, giving a highly variable X-ray power law emission \citep{1993ARA&A..31..717M}.

In this moderate $\dot{m}$ regime, the AGN structure resembles most the one depicted in the geometrical unified model \citep{1993ARA&A..31..473A}, with a well-formed obscuring torus and a BLR, both reacting to the strong photoionizing AGN continuum emission and producing reprocessing features due to scattering, reflection, and absorption, in the AGN spectra both in the optical/UV and X-ray band.
The polar jet emission is suppressed compared to the previous $\dot{m}$ regimes, and the AGN feedback is mainly in radiative form.
The radiation produced in the AGN central engine can additionally significantly affect the accretion flow by depositing enough momentum to drive wide-angle, subrelativistic mass outflows.
In particular, the low-ionization BLR can be explained by a failed continuum radiation-driven dusty wind on radial scales of the order of the dust sublimation radius \citep{2017ApJ...846..154C}, that is, a torus wind.
The high-ionization BLR can instead be explained by an inner failed LD disk wind:
as a cold accretion disk is formed and the AGN is actively accreting, a LD disk wind can be launched \citep[][PK04]{1995ApJ...451..498M,2000ApJ...543..686P}.
However, the energetic output of AGN in this moderate $\dot{m}$ regime is too large in the X-ray band compared to the UV band to allow for a persistent LD disk wind to be formed \citep{1995ApJ...454L.105M}, and there is a large radial zone where the matter gets overionized by the strong central continuum emission, fails to become part of the wind, and forms a failed accretion disk wind \citep{2005ApJ...630L...9P}.
We identify this (mainly failed) wind component with the high-ionization BLR.
Only at outer radii, where the strong ionizing radiation is filtered by the failed inner wind, can LD disk winds be launched, reaching terminal velocities on the order of a few $1000$ km s$^{-1}$, which would manifest as low-velocity X-ray or UV NALs and mini-BALs.
The BH mass is important in this $\dot{m}$ regime, as described in Sect. \ref{SEC:BHMASS}. In particular, for $M_{BH}\gtrsim 10^8 M_{\odot}$ the disk is cold enough to provide a sufficient number of UV photons to push a wind from accretion disk scales, therefore explaining somewhat broader, deeper, and moderate-velocity ($5000$ km s$^{-1}\lesssim \upsilon_{out}\lesssim 0.1c$ ) features such as mini-BALs and low-velocity UFOs.

Strong reprocessing features are imprinted in the optical, UV, and X-ray spectra of AGN in this $\dot{m}$ regime, both in terms of emission and reflection, and blueshifted absorption when the wind features intercept the line of sight \citep[hereafter l.o.s.; see, e.g.,][]{2010MNRAS.408.1396S,2016MNRAS.458..293M}.
When the l.o.s. intercepts the wind, absorption features due to gas with a much larger column density ($N_H\sim 10^{21-23}$ cm$^{-2}$) compared to the weak features due to low-density outer winds of the previous $\dot{m}$ regimes, will appear in the X-ray and UV spectra.

\subsection{High $\dot{m}\gtrsim 0.25$\label{SEC:H}: Wind-dominated active galactic nuclei}
In this $\dot{m}$ regime, powerful radiation-driven accretion disk winds are launched.
The optically thick, geometrically thin accretion disk inner radius extends very close or down to ISCO, and its optical and UV radiative output dominates the AGN SED over the X-ray emission due to the corona, which is significantly cooled down by the interaction with the disk itself.
Above $\dot{m}\gtrsim 0.25$, the radiation pressure is large enough and the relative contribution of X-ray photons over the UV ones is low enough, to allow the production of strong, persistent LD disk winds avoiding strong overionization \citep[PK04,][]{2010A&A...516A..89R}.
The formation of the most powerful LD disk winds is, however, challenging in very low BH mass AGN, as discussed in Sect. \ref{SEC:BHMASS}: in this case the absorption features will resemble more mini-BALs than BALs.
On the contrary, for very large BH masses the disk temperature is low enough to allow the launch and acceleration of powerful (massive, $N_H\sim 10^{22-24}$ cm$^{-2}$, fast, persistent) accretion disk winds that can develop over a vast radial range, and thus producing BALs when observed along the l.o.s..

\subsection{Very high $\dot{m} \gg 1$\label{SEC:VH}: Super-Eddington sources}
In this $\dot{m}$ regime, the AGN is completely outflow dominated, presenting strong, radiation driven disk winds and polar jets \citep{2014MNRAS.441.3177M,2014ApJ...796..106J}.
The geometry and the physics of the inner accretion flow deviate from a geometrically thin and optically thick accretion disk: the inner part of the disk becomes radiation pressure dominated, puffs up, and becomes geometrically thick \citep[slim disk,][]{1988ApJ...332..646A}. The inner slim disk obscures the very inner accretion flow and favors the development of polar jets; there, the emitted radiation can be beamed  by factors of $2-10$ \citep{2016MNRAS.456.3929S}.
The radiative efficiency is uncertain, from a few thousandth \citep{2014MNRAS.439..503S} to a few percent \citep{2017arXiv170902845J}; in any case at least one order of magnitude below the radiative efficiency of luminous AGN accreting at moderate or high $\dot{m}$. In this regime, the inner hot plasma will be the coolest and densest among the five different $\dot{m}$ regimes, and the X-rays will be intrinsically the weakest when compared to the optical and UV emission from the disk, which dominate the SED.
The strongest absorption features will be observed when the l.o.s. intercepts the winds, with terminal velocities of tenths of $c$ and column densities $N_H > 10^{23}$ cm$^{-2}$.

\section{Comparison with observations}\label{OBS}

Active galactic nuclei display a plethora of different observational properties that can be explained to a large extent in the scenario outlined in Sect.~\ref{MAIN}, where the inner accretion/ejection flow changes with  $\dot{m}$ and $M_{BH}$.

\subsection{Broad, mini-broad, and narrow absorption lines}

Broad absorption line, mini-broad absorption line, and narrow absorption line (BAL, mini-BAL, and NAL) QSOs are AGN that host intrinsic blueshifted UV absorption lines of decreasing width, indicating the presence of massive winds coming from their inner regions, with a decreasing range of velocities observed along the l.o.s..
Among these, BAL QSOs were the first to be discovered \citep{1967ApJ...147..396L}, and are by far the best-studied sources.
BAL features are observed in about 10-15\% of optically selected QSOs; the observed fraction of BAL QSOs has to be corrected for absorption and selection effects, to an intrinsic fraction of about 20-25\% \citep[e.g.,][]{1991ApJ...373...23W, 2009ApJ...692..758G}.
In the IR band, the intrinsic fraction of AGN hosting BALs is larger than 40\% \citep{2008ApJ...672..108D}, indicating strong selection effects against the detection of BALs in the optical.
It has been debated for years whether the intrinsic fraction of BAL QSOs corresponds to the covering fraction of a wind always present in all AGN \citep[e.g.,][]{1995ApJ...451..498M}, or to the duty cycle of a fully covering wind present only during a phase of life of the AGN \citep[e.g.,][]{1993ApJ...413...95V}.
In the scenario presented in this work, the answer is between these two: as the presence and strength of the LD disk wind depend on the $\dot{m}$ and $M_{BH}$ of the AGN, the intrinsic fraction of  BAL QSOs contains information on both the AGN wind duty cycle, and the wind geometrical covering fraction.
To disentangle these two physical concurring factors from the dependence on $\dot{m}$ and $M_{BH}$ is the task of ongoing theoretical and observational efforts.
The percentage of AGN hosting mini-BALs and NALs is more uncertain, however it is estimated that all together, BAL, mini-BAL, and NAL features are present in more than 60\% of all QSOs \citep{2008ApJ...672..102G,2012ASPC..460...47H}.
These large fractions demonstrate that the investigations of BAL, mini-BAL, and NAL QSOs are crucial to understand the geometry and the physics of the inner regions of AGN in general.

Broad absorption line QSOs appear X-ray weak with respect to non-BAL QSOs \citep{1995ApJ...450...51G}.
On one hand, such X-ray weakness is consistent with being due to X-ray absorption \citep{2000ApJ...528..637B,2001ApJ...546..795G}, and the intrinsic SED of BAL QSOs with being typical of unabsorbed, non-BAL AGN \citep{2007ApJ...665..157G}. On the other hand, a significant fraction of X-ray bright, X-ray unabsorbed BAL QSOs exists, albeit with less powerful outflows than the X-ray weak ones \citep{2008A&A...491..425G}.
The strength\footnote{The strength of the blueshifted UV absorption can be quantified by means of the Balnicity Index \citep[BI,][]{1991ApJ...373...23W}, the Absorption Index \citep[AI,][]{2002ApJS..141..267H}, or modifications of a generic index of absorption (\citet{2018A&A...617A.118S}). All these indices are integrals of the area subtended by the absorption troughs and are a measure of their equivalent width, blueshift, and width; these physically correspond to the column density and covering fraction, terminal velocity, and velocity dispersion/range of velocities of the winds producing the absorption troughs.} of the blueshifted UV absorption increases with the UV luminosity \citep{2008ApJ...672..102G}, as well as with the X-ray weakness of the AGN \citep{2002ApJ...569..641L,2018MNRAS.479.5335V}, and with the amount of X-ray absorption \citep{2008A&A...491..425G}, as predicted in LD disk winds scenarios. Recent hard X-ray observations performed with  the Nuclear Spectroscopic Telescope Array (NuSTAR) revealed the intrinsically X-ray weak character of powerful BAL QSOs that are also strongly X-ray absorbed \citep{2013ApJ...772..153L,2014ApJ...794...70L}. The strongest BAL QSOs thus belong to the high and very high $\dot{m}$ regimes, where the optical/UV emission from the disk is dominant over the X-ray coronal emission, and the strongest LD disk winds are expected.
In the moderate $\dot{m}$ regime (or in the high $\dot{m}$ regime with small $M_{BH}\ll 10^8 M_\odot$), the UV absorption features would be weaker, such as mini-BALs and NALs, and hosted in relatively X-ray louder AGN, as observed \citep{2008A&A...491..425G,2009ApJ...696..924G,2009NewAR..53..128C}.

The physical relation between NAL, mini-BAL, and BAL QSOs is not yet understood.
The observed UV/X-ray properties of mini-BAL QSOs are in between those of BAL and non-BAL  QSOs \citep{2009ApJ...696..924G,2010ApJ...724..762W}, and
the X-ray weakness increases going from NAL to  mini-BAL to BAL QSOs \citep{2009NewAR..53..128C}. This could mean a geometrical difference of our l.o.s., where BAL, mini-BAL, and NAL structures may coexist in the same AGN while occupying higher and higher latitudes above the accretion disk plane \citep{2012ASPC..460...47H}; or a physical difference, with, for example,   mini-BALs being seeds capable of evolving into BALs if the AGN spectral energy distribution changes favorably \citep{2009ApJ...696..924G}.
In  recent years, there have  been observations of  non-BAL QSOs developing BALs \citep{2008MNRAS.391L..39H, 2010ApJ...724L.203K}, of NALs evolving
into BALs \citep{2002MNRAS.335L..99M}, of mini-BALs evolving into BALs
\citep{2013ApJ...775...14R}; the canonical Seyfert1  NGC 5548 has been observed in  an X-ray obscured state, and has been discovered to have concurrently developed UV mini-BALs \citep{2014Sci...345...64K}.
In some cases, the difference between BALs and mini-BALs and NALs could also be due to instrumental effects, that is, to a spectral resolution insufficient to disentangle multiple, narrow absorption troughs that therefore appear as a unique, artificially broader absorption trough \citep{2018MNRAS.474.3397L}.
All in all, the BAL-nicity of an AGN will depend on both the physical configuration of the wind intercepted by our l.o.s., and on the geometrical angle of inclination between our l.o.s. and the wind itself: in our picture, all Seyfert galaxies will appear as NAL or mini-BAL QSOs when observed along the ``right'' l.o.s, and all QSOs will appear as mini-BAL or BAL QSOs when observed along the ``right'' l.o.s..

A value of the Absorption Index AI~$\sim 1000-2000$ km s$^{-1}$ is found to separate the strongest BAL features, which correspond to highly X-ray absorbed AGN, from the weaker ones, which correspond to lowly X-ray absorbed and unabsorbed AGN \citep{2008MNRAS.386.1426K, 2008A&A...491..425G}.
We suggest that the AI is inversely proportional to the launching distance of the wind, and that the AI~$\sim 1000-2000$ km s$^{-1}$ observed threshold corresponds to a physical difference: larger AI values correspond to UV-absorbing features due to stronger winds that are launched by either radiation or magnetic pressure from small scales in the accretion disk; while lower AI values correspond to features due to weaker winds, which are launched in the outer parts of the accretion disk or even farther away from the central SMBH, and that could be easily accelerated by thermal pressure or by large-scale shocks due to the impact of the innermost faster winds on the intergalactic medium \citep[e.g.,][]{2012MNRAS.420.1347F}.
This could be the case of the  BALs that are inferred to be situated at more than 100 pc from the central SMBH \citep{2018ApJ...857...60A}.

\subsection{Presence or absence of line-driven disk winds and the active galactic nuclei structure}

 \begin{figure*}[h]
\centering
\includegraphics[width=16cm]{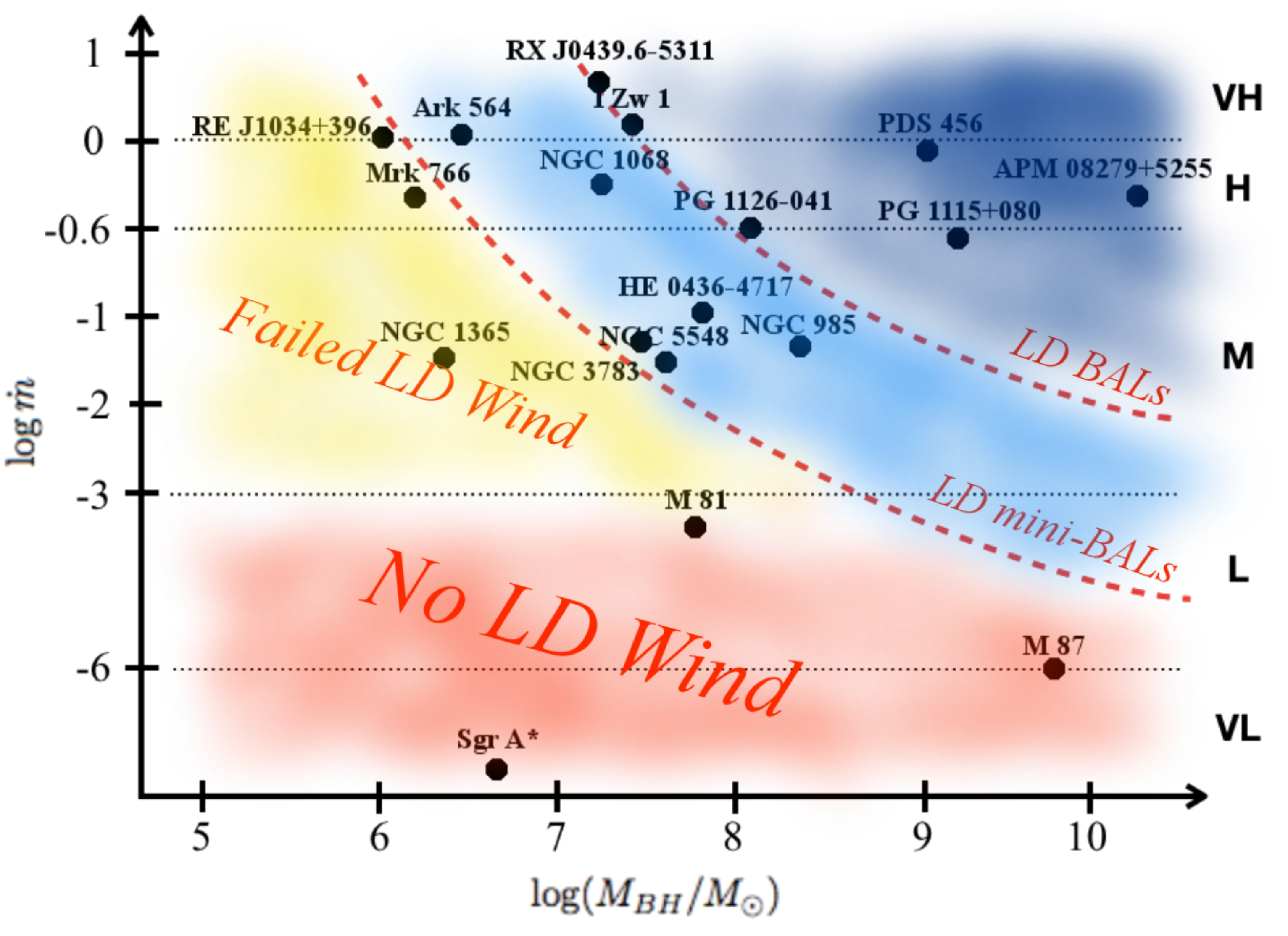}
   \caption{Eddington ratio $\dot{m}$ and black hole mass $M_{BH}$ parameter space with the five different $\dot{m}$  regimes marked by dotted lines (note that the vertically logarithmic scale is nonlinear). The two dashed curves mark approximately the lower limit in the $\dot{m}-M_{BH}$ space for the existence of UV LD mini-BALs (bottom curve) and BALs (upper curve).
   The location of various AGN mentioned in this article is marked in the figure.
   References for the individual measurements of $\log(M_{BH}/M_{\odot})$ and $\dot{m}$ (or $L_{BOL}$, assuming $\eta=0.1$ for luminous AGN) are taken from the literature and reported in parentheses, respectively from: \citet{2004AJ....127.3168B} and \citet{2017MNRAS.468.3489K} for Ark 564 ($6.4,1.1$);  \citet{2006ApJ...641..689V} and \citet{2017MNRAS.468....2M} for I Zw 1 ($7.4, 1.4$); \citet{2009ApJ...705..199B} and \citet{2010MNRAS.402.1081V} for Mrk 766 ($6.2,0.4^*$); \citet{1996ApJ...472L..21G}  for NGC 1068 ($7.2,0.5$); \citet{2013Natur.494..449R} and \citet{2010MNRAS.402.1081V} for NGC 1365 ($6.3,0.025$); \citet{2014MNRAS.445.3073P} and \citet{2010MNRAS.402.1081V} for NGC 5548 ($7.6,0.025$); \citet{2010ApJS..187...64G} and \citet{2018A&A...618A.167M} for HE 0436-4717 ($7.8,0.09$); \citet{2007ApJ...657..102D} and \citet{2011A&A...536A..49G} for PG 1126-041 ($8.1,0.25$); \citet{2006ApJ...649..616P} and \citet{2007AJ....133.1849C} for PG 1115+080 ($9.1,0.2$); \citet{2018A&A...617A.118S} for APM 08279+5255 ($10.4,0.4$); \citet{2015Sci...347..860N} for PDS 456 ($9,0.8$); \citet{2017MNRAS.471..706J}  for RX J0439.6-5311 ($7.2,1.8$), \citet{2016A&A...594A.102C} for RE J1034+396 ($6,1$); \citet{2010MNRAS.402.1081V} for NGC 985 ($8.4,0.05$); \citet{2010MNRAS.402.1081V}  for NGC 3783 ($7.5,0.05$); \citet{2003AJ....125.1226D} and \citet{1996ApJ...462..183H} for M81 ($7.8,6\times 10^{-4}$); \citet{2009ApJ...700.1690G}\citet{2016MNRAS.457.3801P} for M87 ($9.8,10^{-6}$); and \citet{2008ApJ...689.1044G} and \citet{2009ApJ...706..497M} for Sgr A* ($6.6,<10^{-6}$).\\(*) \citet{2010MNRAS.402.1081V} estimate a lower $\dot{m}$ for Mrk 766 as they use a larger BH mass than the one used here.}    \label{bigscheme}%
    \end{figure*}

In Fig.~\ref{bigscheme} we plot the locations of several AGN, which are mentioned in this article, in the $\dot{m}-M_{BH}$ plane, where the $M_{BH}$ and $\dot{m}$ values are taken from the literature. The five $\dot{m}$ regimes introduced in our  scenario are separated by horizontal dotted lines.
 Two further red dashed lines represent the approximate lower boundary in terms of $\dot{m}$ and $M_{BH}$ for the expected presence of LD mini-BALs and BALs. The lines delimiting the various areas are rather fuzzy to represent the divisions between the regimes, which are not  rigidly defined\footnote{At the boundary between different $\dot{m}-M_{BH}$ regimes there is no sudden switch between the presence or absence of, for example, mini-BALs, but a smooth transition of physical properties between systems capable of, for example, launching strong mini-BALs toward systems able to launch BALs, with some sources indeed presenting a superposition of the two phenomena, or a transition in time between the two phenomena \citep[e.g.,][]{2013ApJ...775...14R, 2017MNRAS.468.4539M}.
 A more precise identification of the location of the transition between mini-BALs and BALs in the $\dot{m}-M_{BH}$ parameter space, as well as the inclusion of large samples of AGN at different redshift into the $\dot{m}-M_{BH}$ plane as a function of their intrinsic UV/X-ray SED and outflow properties, are tasks deserving of  future investigations.}.

At very low and low $\dot{m}$, the accretion flow is hot and geometrically thick, and favors the development of relativistic polar jets, whose emission dominates the SED \citep[e.g., Sgr A*, see][]{2013A&A...559L...3M}.
In particular, at very low $\dot{m}$,  LD disk winds cannot develop at any $M_{BH}$, due to the absence of an accretion disk.
At low $\dot{m}$ instead, an outer accretion disk can be present, but truncated at large radii ($\sim 10^3-10^4 R_g$), as inferred, for example, from the detailed X-ray spectral analysis of the nucleus of the LLAGN in NGC 4258 \citep{2009ApJ...691.1159R}, or from the broadband SED fitting of a sample of LLAGN \citep{2014MNRAS.438.2804N}.
The LLAGN will appear as ``true type 2'' AGN, with very low intrinsic UV and X-ray obscuration, as there is no or little disk, no or little BLR, and no or little equatorial torus, as observed by \citet{1999ApJ...516..672H,2012ApJ...748..130M,2014MNRAS.438.2804N}.
The presence of an outer disk might allow the launch of LD disk winds, but only in the (relatively) higher $\dot{m}$ and largest $M_{BH}$ systems. These winds will be launched at relatively large distances from the SMBH and over a narrow range of radii, therefore observationally will produce weak, low-ionization BLR emission lines, and low column density, low-velocity ($\upsilon_{out}\ll 2000 $ km s$^{-1}$) mini-BALs in absorption.

At moderate and high $\dot{m}$, LD disk winds can be launched.
Here  the AGN are in the Seyfert or QSO regime and resemble most the model of \citet{1993ARA&A..31..473A}: a torus, a disk, and the BLR are indeed predicted \citep{2000ApJ...530L..65N,2009ApJ...701L..91E} and observed to exist \citep{1994ApJS...95....1E,2011ApJS..196....2S,2015ARA&A..53..365N}. These introduce a fundamental geometrical difference in observational properties between sources seen at large inclination angles (obscured, type 2 AGN), and sources observed at low inclination angles (unobscured, type 1 AGN).
In the scenario outlined in this work, the differences between type 1 and type 2 AGN will, however, be not only geometrical, but also physical, due to the presence or absence of strong LD disk winds in different $\dot{m}$ and $M_{BH}$ regimes.
The covering factor of the LD disk wind will crucially depend on the X-ray/UV flux ratio, and will increase with increasing $\dot{m}$ and $M_{BH}$:
for low $M_{BH}$ and $\dot{m}$, the wind will be mainly failed, while increasing $M_{BH}$ and $\dot{m}$ will increase the effectiveness of radiation pressure to launch fast winds over a large disk radial range.
Therefore for the observer, the AGN will appear as Seyfert, NLS1s/mini-BALs, and BAL QSOs with increasing $\dot{m}$ and $M_{BH}$.
One prediction of this scenario is therefore that Seyfert galaxies with large enough $M_{BH}$ and $\dot{m}$ will develop mini-BALs; and these will be less sporadic, broader, deeper, and faster as $M_{BH}$ and $\dot{m}$ increase. The presence of time-varying mini-BALs with outflow velocities up to $\sim 5000 $ km s$^{-1}$ has been revealed by high-quality spectroscopic observations of the moderately accreting Seyfert 1 galaxies NGC 3783, NGC 5548, and NGC 985, with their strength being inversely proportional to the soft X-ray flux of the AGN \citep{2017A&A...607A..28M, 2014Sci...345...64K, 2016A&A...586A..72E}, as predicted in LD disk winds scenarios.

By increasing $M_{BH}$ and $\dot{m}$, the AGN will eventually enter the BAL QSO regime, where powerful, wide-angle LD disk winds are launched \citep[e.g., APM 08279+5255, PDS 465; see, e.g.,][]{2009ApJ...706..644C, 2015Sci...347..860N}.
The effects of radiation pressure create the BLR: the bulk of the low-ionization BLR (e.g., H$\beta$, Mg \textsc{ii}) is created by continuum radiation pressure on dusty gas at torus (i.e., outer disk) scales, and reacts to continuum luminosity changes on timescales of days; while the bulk of the high-ionization BLR (e.g., C \textsc{iv}, N \textsc{v}) is created by spectral line radiation pressure acting on sub-parsec scales, that is, on the inner accretion disk, and varies on shorter timescales \citep{2000ARA&A..38..521S,2004ApJ...613..682P,2016ApJ...821...56F}. Furthermore, part of the low-ionization BLR can be formed within the LD disk wind, either in cold clumps within the wind, or at the wind base itself.
It has been shown by \citet{2013ApJ...778...50K} that at the base of a slowly accelerating wind that originates in a Keplerian disk, the virialization of the flow is maintained, contrary to the main body of the wind that will be significantly outflowing.
Therefore the base of the LD disk wind can contribute to the symmetric, slowly blueshifted low-ionization BLR, while the bulk of the wind material itself can contribute to the asymmetric, strongly blueshifted high-ionization BLR \citep{2011AJ....141..167R}.
As the wind launching region successfully extends over a vast radial portion of the inner flow at the expense of a less extended zone of failed wind, at high $\dot{m}$ reprocessing features such as the BLR are still present in the SED albeit with relatively less prominence than in the moderate $\dot{m}$ case \citep[thus qualitatively explaining the optical/UV and X-ray Baldwin/Iwasawa-Taniguchi effects,][]{1977ApJ...214..679B,1993ApJ...413L..15I}, and show evident effects of reprocessing into the wind, for example, strong blueshift or asymmetry  \citep[e.g.][]{2008MNRAS.388..611S}.

At very high $\dot{m}>1$, the system is outflow dominated both in the polar and the equatorial region: the strongest UV and X-ray absorbing winds are produced, and the covering factor of the wind will be maximum, not far from $4\pi$ sr. Low redshift AGN observed in this regime are often extreme NLS1 where the SED is disk dominated, for example RE J1034+396, RX J0439.6-5311 \citep[e.g.,][]{2009MNRAS.398L..16J,2017MNRAS.468.3663J}. At higher redshift, probing larger BH masses and more luminous AGN in the QSO regime, the inner accretion flow of super-Eddington sources is inferred to be self-obscured, giving X-ray weak sources with a very weak high-ionization BLR and reprocessing features \citep[e.g., PHL 1811 and analogs,][]{2015ApJ...805..122L}. Recent observations of luminous QSOs with a very weak BLR revealed indeed very steep X-ray spectra, indicative of high $\dot{m}$ \citep{2018ApJ...865...92M}.

From a general point of view, at very low and low $\dot{m,}$ the accretion and the ejection flows are almost isotropic, that is, the inclination angle of the observer's l.o.s. is generally unimportant, except for the polar region, where the relativistic jet emission is observed being boosted when observed parallel to the l.o.s. \citep[e.g.,][]{1993ApJ...407...65G}. On the contrary, at moderate and high $\dot{m}$, the accretion and ejection flows are more equatorial, and the presence of LD accretion disk winds and their failed component makes the accretion and ejection flow very structured: therefore the appearance of the AGN will depend a lot on the inclination of the observer's l.o.s..
At very high $\dot{m}$, the presence of both equatorial and polar outflows makes the accretion and ejection flow again quite isotropic.
The geometrical effects dependent on the observer's l.o.s. for the moderate and high $\dot{m}$ AGN are discussed in Sect. \ref{SEC:GEOM}.

\subsection{Geometrical effects due to the presence of line-driven disk winds}\label{SEC:GEOM}

\begin{figure*}[h]
\centering
\includegraphics[width=13cm]{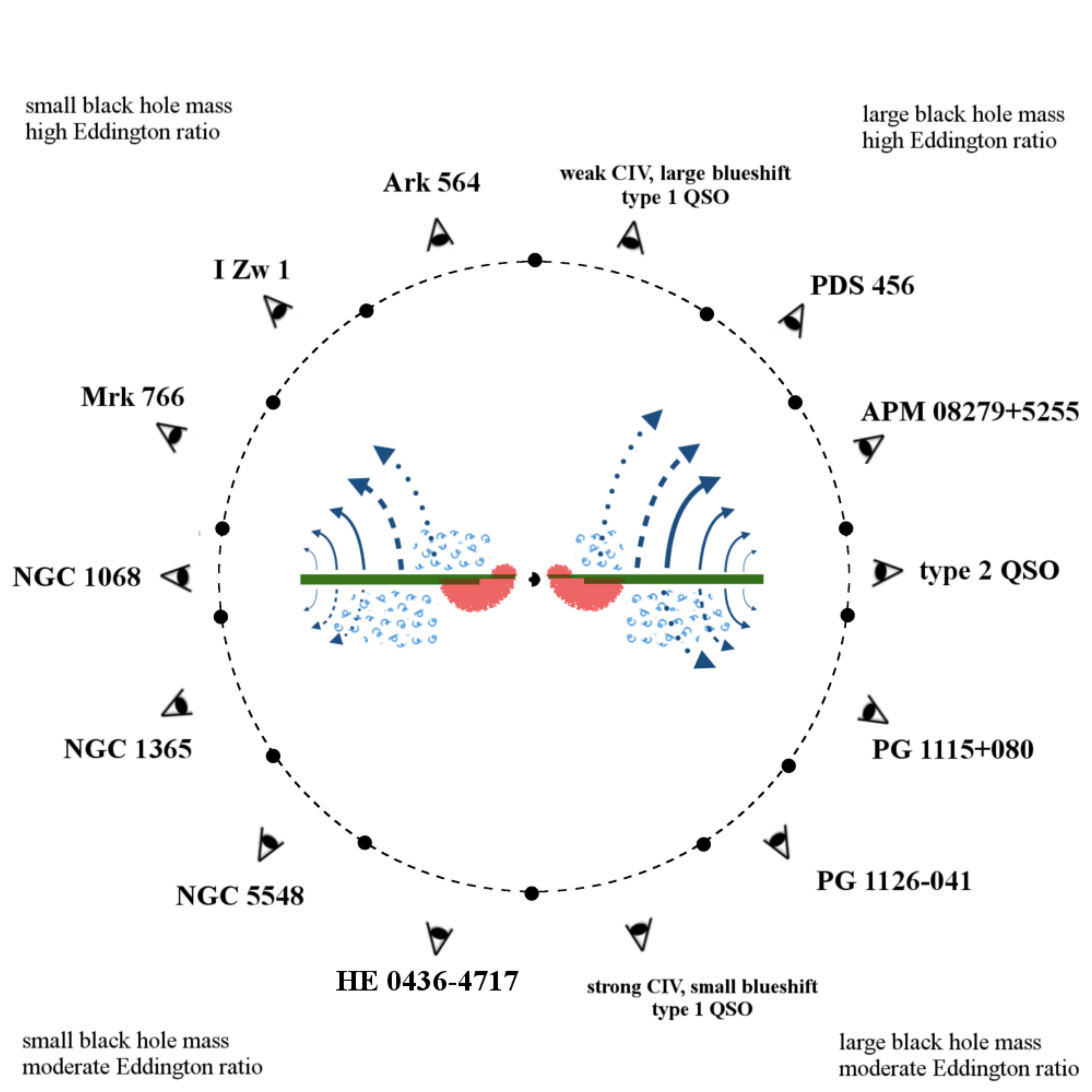}
   \caption{Sketch of the inner accretion/ejection flow for luminous AGN in four $\dot{m}, M_{BH}$ regimes, and different lines of sight labeled together with names of known AGN. Bottom left: $M_{BH}\ll 10^8 M_{\odot}, \dot{m}\approx 10^{-3}-10^{-1}$; top left: $M_{BH}\ll 10^8 M_{\odot}, \dot{m}\gtrsim 0.25$; top right: $M_{BH}\gtrsim 10^8 M_{\odot}, \dot{m}\gtrsim 0.25$; bottom right: $M_{BH}\gtrsim 10^8 M_{\odot}, \dot{m}\approx 10^{-3}-10^{-1}$.  }    \label{FIGURAlos}%
    \end{figure*}

 The presence or absence of strong equatorial disk winds introduces an important element of anisotropy for the AGN accreting at moderate and high $\dot{m}$. We  illustrate in Fig. \ref{FIGURAlos} different l.o.s. toward the central engine of AGN in these $\dot{m}$ regimes, also differentiating between the small and large $M_{BH}$ cases.
 Only for polar or quasi-polar l.o.s. will the AGN central engine be clearly visible.
 In this case the AGN will appear ``bare'', with a minimum amount of X-ray and UV absorption (with column densities $N_H\lesssim 10^{20-21}$ cm$^{-2}$), as observed in local type  Seyferts such as Ark 120 \citep{2004MNRAS.351..193V} or Ark 564 \citep{2015A&A...577A...8G}; or in general, as observed in luminous, unabsorbed, type 1 QSOs.
 However, the inner environment of such AGN will not be completely bare, as plenty of gas is still present outside the l.o.s., as demonstrated by the reprocessing features detected in long X-ray observations of Ark 120 \citep{2016ApJ...828...98R}, or from the strongly blueshifted high-ionization broad emission lines observed in the spectra of luminous type 1 QSOs, indicating reprocessing in large columns of outflowing gas \citep{2011AJ....141..167R}.

 For non-polar l.o.s., the LD disk wind will be intercepted, and the intervening material will introduce significant reprocessing features in the observed SED \citep[e.g.,][]{2009ApJ...694....1S,2010MNRAS.408.1396S}.
In particular, the observed absorption features due to the wind material will be weak or strong, depending both on how intrinsically strong the wind is, and on how deep into the flow the l.o.s. is going.
Specifically, broad absorption lines (BALs) are produced when a wide range of velocities are involved in the flow as seen by the observer, either because of an intrinsically wide spread of velocities into the flow, or because of inclination angle effects \citep[e.g.,][]{2012ApJ...758...70G}.
 Narrower absorption features (mini-BALs) will be observed in AGN with intrinsically narrower radial ranges where a disk wind can be launched, that is, in moderate $\dot{m}$, small $M_{BH}$, compared to high $\dot{m}$, large $M_{BH}$ AGN. Also for high inclination angles through an equatorial wind, where mainly the transverse component of the velocity vector is observed along the l.o.s., narrower absorption features will be observed.

The low $M_{BH}$, moderate $\dot{m}$ case (bottom left quadrant in Fig. \ref{FIGURAlos}) and the large $M_{BH}$, high $\dot{m}$ case (top right quadrant in Fig. \ref{FIGURAlos}) present very different wind configurations that are easily distinguishable. In the former  case, there are only low-velocity radiation-driven winds with sporadic mini-BALs launched at large radii, and there is a vast radial range of failed wind \citep[e.g.,][]{1995ApJ...454L.105M,2005ApJ...630L...9P}; this physical regime will correspond to Seyfert galaxies.
In the latter case, the ejection flow is dominated by radiation pressure, with well-developed LD disk winds originating from a vast range of disk radii (PK04); this physical regime will correspond to BAL QSOs.
It is instead difficult to distinguish the low $M_{BH}$, high $\dot{m}$ case (top left quadrant in Fig. \ref{FIGURAlos}) and the large $M_{BH}$, moderate $\dot{m}$ case (bottom right quadrant in Fig. \ref{FIGURAlos}). This is because in both cases their disk winds will appear similar, with mini-BALs and only low-velocity or sporadic BALs. This is due to the joint effect of the higher $\dot{m}$ (favoring a LD disk wind) and the hotter disk (disfavoring a LD disk wind) in the former case, compared to the latter. While the phenomenology of Seyfert galaxies and BAL QSOs is very different, the observed characteristics of NLS1s and mini-BALs can be present in the same source \citep[e.g., PG 1126-041 and Mrk 335, see respectively][]{2011A&A...536A..49G,2013ApJ...766..104L}.

For AGN with small $M_{BH}\ll 10^8 M_{\odot}$ in the moderate $\dot{m}$ regime, episodic obscuration and outflows with low velocity ($\upsilon_{out}< 2000$ km s$^{-1}$) and very narrow (FWHM $\ll 2000$ km s$^{-1}$) UV absorbing mini-BALs features can be observed at moderate inclination angles of the l.o.s. \citep[e.g., NGC 5548,][]{2014Sci...345...64K}. At more equatorial inclination angles, plenty of the failed wind component will be intercepted, giving less sporadic but still low-velocity UV and X-ray absorption, and UV and X-ray eclipses of the continuum source due to clumps of the failed wind intercepting the l.o.s. \citep[e.g., NGC 1365,][]{2005ApJ...623L..93R}.
  In the high $\dot{m}$ regime, somewhat broader mini-BAL-like winds are feasible and moderate velocity ($\upsilon_{out}\sim 2-5000$ km s$^{-1}$) UV and X-ray absorbing outflows are observed at moderate inclination angles of the l.o.s. \citep[e.g., I Zw 1,][]{1997ApJ...489..656L,2001ApJ...557....2C,2018MNRAS.480.2334S}; while more equatorial l.o.s. will give again X-ray and UV eclipses of the central source \citep[e.g., Mrk 766,][]{2011MNRAS.410.1027R}.

    Stronger X-ray and UV absorption features are expected for AGN with large $M_{BH}\gtrsim 10^8 M_{\odot}$, where a stronger LD disk wind can be launched (PK04), giving ``complex'' and variable spectra when intercepting the more clumpy portion of the well-developed wind.
    Moderate velocity UV and X-ray absorbing features will be observed at intermediate l.o.s. inclination angles in large $M_{BH}$, moderate $\dot{m}$ AGN \citep[e.g., $\upsilon_{out}\sim 0.02-0.05c$ and $N_H\sim 10^{23}$ cm$^{-2}$ as observed in the mini-BAL QSOs of the PG catalog,][]{2016AN....337..459G}, and higher velocity and stronger features ($\upsilon_{out}\sim 0.2-0.3c$ and $N_H\sim$ several $10^{23}$ cm$^{-2}$ ) will be observed for the same inclination angles in AGN in a high $\dot{m}$ regime \citep[as observed for example in PDS 456,][]{2014ApJ...784...77G,2016MNRAS.458.1311M,2018MNRAS.476..943H}.
    More equatorial l.o.s. go deep into the well-developed LD disk wind in large $M_{BH}$ AGN, and the strongest features in terms of width, depth, and blueshift are observed: this will be the case of the moderate $\dot{m}$ mini-BAL QSOs \citep[e.g., $\upsilon_{out}\sim 0.1-0.3c$ and $N_H\sim 5 \times 10^{22}-5 \times 10^{23}$ cm$^{-2}$ as observed in  PG 1115+080,][]{2003ApJ...595...85C} and the high $\dot{m}$ BAL QSOs  \citep[e.g., $\upsilon_{out}\gtrsim 0.4c$ and $N_H\sim$ several $10^{23}$ cm$^{-2}$ as observed in APM 08279+5255,][]{2002ApJ...579..169C,2009ApJ...706..644C,2011ApJ...737...91S}.

The absorption will become stronger and stronger going toward equatorial l.o.s., until reaching the Compton-thick regime ($N_H\gtrsim 10^{24}$ cm$^{-2}$) in type 2 AGN such as NGC 1068 or generally type 2, absorbed QSOs, where the primary emission due to the central engine is totally suppressed and only hard X-ray emission and reprocessing features are observed \citep{1997A&A...325L..13M,2002ApJ...575..732K,2015ApJ...812..116B}.
 The circumnuclear absorbers are actually observed to be clumpy on all scales \citep{2014MNRAS.439.1403M,2017NatAs...1..679R}, therefore there is still a non-null probability of observing the AGN central engine also in obscured, type 2 AGN, in case of occultation events caught on diverse scales, like on the accretion disk and BLR-scale \citep[e.g.,][]{2005ApJ...623L..93R,2010A&A...517A..47M}, or on the torus-scale \citep[e.g.,][]{2011ApJ...742L..29R,2015ApJ...815...55R,2016MNRAS.456L..94M}.

\section{Discussion}\label{DISC}

We propose a qualitative global scenario for AGN that includes different accretion/ejection mechanisms depending on $\dot{m}$ and $M_{BH}$.
We assume that with increasing $\dot{m}$, a central hot flow shrinks while an outer cold flow extends more and more toward the central SMBH, similar to the models of \citet{2000A&A...360.1170R} and \citet{2004A&A...414..895F}.
At the lowest Eddington ratios (very low and low $\dot{m}$ in our nomenclature),  the geometrical and physical configurations of the AGN are comparable to the ones described by these authors, while at moderate and high $\dot{m}$, LD disk winds and their inner failed component are introduced to explain the diverse appearance of the AGN inner structure.
A popular model aimed at unifying the AGN inner structure through the presence of a wind on accretion disk scales is the one proposed by \citet{2000ApJ...545...63E}. In this model, all the AGN diverse appearance is explained by different inclination angles of the l.o.s. with respect to a hollow wind, which initially rises perpendicular to the disk, and then bends into a biconical outflow. This wind geometry would give no absorption when the l.o.s. is polar, deep and strongly blueshifted BALs when the l.o.s. is intermediate and goes through the flow, and narrower absorption features such as X-ray warm absorbers and UV mini-BALs and NALs when the AGN is observed more perpendicularly across the flow.
Contrary to the \citet{2000ApJ...545...63E} model, where the wind is geometrically thin, here the accretion disk wind is extended over a large range of radial distances from the central SMBH. Furthermore, in our scheme radiation-driven accretion disk winds are not always present. They are absent in the very low and low $\dot{m}$ regimes, and present but weak in the case of moderate $\dot{m}$, where the wind is mainly failed; in the high and very high $\dot{m}$ regimes the wind has a large duty cycle and geometrical covering fraction, with a less relevant failed component. The geometrical covering fraction of the LD disk wind will thus increase with increasing $\dot{m}$ and $M_{BH}$.
This is relevant for the amount of AGN kinetic feedback, which will also be proportional to the LD disk wind strength and will therefore increase going from Seyfert galaxies to QSOs.

The UV/X-ray radiative output from the AGN central engine is crucial for the acceleration of LD disk winds. In the moderate and high $\dot{m}$  regimes, the UV/X-ray flux depends on the relative energetic contribution of the cold, optically thick matter of the UV emitting accretion disk, and the hot, optically thin matter of the inner X-ray emitting corona. The energetic contribution to the SED of the former is observed to increase over the latter with increasing $\dot{m}$ \citep{2009MNRAS.392.1124V,2012MNRAS.425..907J}.
This is explained with the increased inner radial extent of the UV disk, at the expenses of the X-ray corona, at larger $\dot{m}$ compared with lower $\dot{m}$. Assuming a constant radiative efficiency above $\dot{m}\sim 0.01$, the relative UV/X-ray flux ratio will thus increase with increasing $\dot{m}$, moving from low-luminosity Seyfert galaxies toward higher luminosity-Seyfert galaxies and QSOs.
A consequence of our scenario is therefore that the X-ray coronae in Seyfert galaxies will be extended up to a few tens $R_g$, different than in QSOs where they will be very compact, only a few $R_g$ in size.
Recent observations of distant microlensed QSOs revealed very compact X-ray coronae only $\lesssim10 R_g$ in size \citep{2009ApJ...693..174C,2010ApJ...709..278D,2015ApJ...806..258M}. In comparison, deep and broadband X-ray spectroscopic observations of the local Seyfert Ark 120 suggest an extended corona \citep[$\sim 10-40 R_g$,][]{2014MNRAS.439.3016M}, and a decrease of the corona size with increasing flux of the continuum source \citep{2019A&A...623A..11P}.
Studies of larger samples of sources are clearly needed in order to assess the presence of a statistical difference in X-ray coronal size between Seyfert galaxies and QSOs.

Radiation-driven disk winds driven by pressure on spectral lines are a physical component that can explain many observational features of AGN, but not all of them: for example, the BALs observed in the FR I radio galaxies \citep{1974MNRAS.167P..31F}, which are accreting at low $\dot{m}$, must be driven by magnetic pressure.
In our scenario, the low and very low $\dot{m}$ AGN appear as radio loud because they lack strong optical and UV emission, and would correspond to the FR I, or LEG \citep[low excitation galaxies,][]{1994ASPC...54..201L}. At higher $\dot{m}$, radio-loud AGN appear as FR II (or HEG, high excitation galaxies) and do have an accretion disk \citep[e.g.,][]{2001A&A...379L...1G, 2004Sci...306..998G}. The radio loudness associated to their powerful radio jets is probably driven by the efficient extraction of rotational energy from both the disk \citep{1982MNRAS.199..883B} and the black hole \citep{1977MNRAS.179..433B}; however, the exact physical mechanism behind the launching and acceleration of the most powerful radio jets is still under investigation \citep[see][for a recent review]{2016AN....337...73C}.
We note how there are no apparent differences in observed properties such as black hole mass, Eddington ratio, and BLR size between radio-loud and radio-quiet BAL QSOs \citep{2014A&A...569A..87B}, suggesting that the physical difference between them might reside in different black hole spin values. The scarcity of powerful radio-loud FR II BAL QSOs is probably due to the combination of a large black hole mass needed to have a RL AGN \citep{2000ApJ...543L.111L}, which is possibly related to the merger history of the host galaxies, where coherent mergers can increase the value of the BH spin \citep[e.g.,][]{2011MNRAS.410...53F}, and a high or very high $\dot{m}$ to have a radiation driven BAL wind, all in all giving a small region of the $M_{BH}-\dot{m}$ parameter space where the phenomenon of radio-loud BAL QSO can manifest.

Highly ionized, high-velocity X-ray UFOs with $\log\xi\sim 3-6$ erg cm s$^{-1}$ and $\upsilon_{out}\sim 0.1-0.4c$ can be launched from very close to the SMBH \citep{2012MNRAS.422L...1T}, and are therefore probably sensitive to the very high ionizing flux, making them highly variable, and temporally transient \citep[e.g.,][]{2012MNRAS.423..165P}. The physical nature of UFOs is still to be clarified: they are observed in both radio-quiet and radio-loud AGN \citep{2010ApJ...719..700T,2014MNRAS.443.2154T}, and they could be launched either radiatively \citep{2017MNRAS.472L..15M} or magnetically \citep{2014ApJ...780..120F}.
 In particular, radiation driven UFOs can be expected in the polar regions of very high $\dot{m}$ AGN, where baryon-loaded jets with velocities of $\upsilon_{out}\sim 0.3c$ are launched \citep{2016MNRAS.456.3929S}; while magnetically driven UFOs can be expected at the base of radio jets in radio-loud AGN accreting at any $\dot{m}$ regime.
The capability of LD of launching disk winds strongly depends on the ionization state of the gas, and it is maximum at moderate ionization states ($\log\xi\lesssim 2$), where the opacity due to UV spectral lines is also maximum; while it decreases significantly at higher ionization states \citep[e.g.,][]{1994ApJ...432...62A}. However, it has been shown by \citet{2018arXiv181201773D} that, given the typical AGN SED, the effectiveness of LD can be significant also in the moderate X-ray regime ($\log\xi\sim 3$), therefore possibly explaining the launch and acceleration of disk winds producing the most moderate (in terms of ionization state and velocity) UFOs.
Sporadic high-velocity ejections from the failed LD disk wind are also observed in the hydrodynamical simulations performed by PK04, possibly explaining the transience with a long duty cycle of the UFOs observed in moderately accreting Seyfert galaxies such as Mrk 509 \citep{2009A&A...504..401C}.
 Detailed analyses of long X-ray observations of local AGN have shown strong variability of the UFO absorption features on very short timescales, of the order of hours \citep{2011A&A...536A..49G,2014ApJ...784...77G}.
 Also the broad UFO features observed in the high redshift QSO APM 08279+5255 are found to be highly variable in energy and intensity on rest-frame timescales of months, implying a non-constant kinetic energy injection in the surrounding media due to the disk wind \citep{2009ApJ...697..194S,2009ApJ...706..644C}.
 Future observations with the high-resolution X-ray spectrometers onboard the X-ray Imaging and Spectroscopy Mission \citep[XRISM,][]{2018arXiv180706903G} and the Advanced Telescope for High-ENergy Astrophysics \citep[ATHENA,][]{2013arXiv1306.2307N} will allow us to reveal the actual duty cycle of UFOs and therefore to unveil the dynamics of the very inner accretion/ejection flow around SMBHs, and its contribution to the AGN feedback on the surrounding environment.


 \section{Summary and conclusions}\label{CONC}

 We propose a global scheme for the inner regions of AGN aimed at simplifying their understanding in terms of different $\dot{m}$ and, secondarily, different $M_{BH}$, which lead to different accretion/ejection flows around the central SMBH.
   We suggest five $\dot{m}$ regimes where the physics of the AGN accretion/ejection flow is expected to significantly change:
   \begin{itemize}
       \item very low $\dot{m}\ll 10^{-6}$, where radiative cooling is negligible, the accretion flow is completely optically thin and geometrically thick, the ejection flow is magnetically driven and mainly polar, with the feedback being mainly kinetic;
       \item low $\dot{m}\approx 10^{-4}$, where radiative cooling starts to be important in the outer equatorial regions but the accretion flow is still dominated by its hot and geometrically thick component, and the ejection flow is still polar and magnetically driven and the feedback mainly kinetic;
       \item moderate $\dot{m}\gtrsim 10^{-2}$, where radiative cooling is efficient and the accretion flow is mainly equatorial, dominated by an optically thick and geometrically thin accretion disk, but strong radiation -driven disk winds cannot develop, especially for small $M_{BH}$, and the feedback is mainly radiative;
       \item  high $\dot{m}\gtrsim 0.25$, where the accretion flow is disk dominated and radiatively efficient, and for large $M_{BH}$ the ejection flow is dominated by equatorial radiation-driven disk winds that can significantly contribute to feedback via the injection of kinetic energy into the surrounding environment;
       \item very high $\dot{m}\gg 1$, where the inner disk puffs up under strong radiation pressure, and both polar and equatorial outflows contribute to the ejection flow.
   \end{itemize}
 These correspond observationally to quiescent and inactive galaxies; LLAGN; Seyferts and mini-BAL QSOs; NLS1s and BAL QSOs; and super-Eddington sources.
 The strongest LD disk winds will be present in the largest $M_{BH}$, highest $\dot{m}$ AGN; while only moderate LD disk winds are possibly present in smaller $M_{BH}$, moderate $\dot{m}$ AGN;  no LD winds are present in low and very low $\dot{m}$ AGN, independent of the $M_{BH}$.
This is a simple scheme that neglects important aspects of accretion onto SMBHs, such as non-stationarity, which will happen in a realistic context.
Albeit simple, our scenario already provides a scheme with which to compare the observations of AGN, and from which to draw future directions for further modeling and observations.
The divisions between the five identified $\dot{m}$ regimes are still qualitative and more quantitative and observational studies are needed to pin down the more precise location of the changes of physical regimes.
Ongoing research is placing large samples of AGN in the $\dot{m}-M_{BH}$ plane as a function of their outflow and X-ray and UV properties, in order to test the global properties of the scenario and to better constrain the diverse physics at work in the inner regions around SMBHs.

\begin{acknowledgements}
MG acknowledges Spanish public funding through the grant ESP2015-65597-C4-1-R.
We acknowledge support provided by the Chandra award G06-17097X issued by the Chandra X-Ray Observatory Center, which is operated by the Smithsonian Astrophysical Observatory for and on behalf of NASA under contract NAS8-39073.
SRON is supported by the Netherlands Organisation for Scientific Research.
We thank Tim R. Waters for insightful suggestions and comments on the manuscript, Chris Done for inspiring discussions and collaborations during the first phase of creation of this work, and Matteo Guainazzi and Elisa Costantini for a careful reading of an early version of the manuscript.
We also thank the referee for providing feedback that helped improve the quality of this work.
MG would also like to thank Massimo Cappi, George Chartas, Gabriele Ponti, Cristian Vignali, Mike Eracleous, Sara Elisa Motta, and Massimo Dotti for either continuous or sporadic, but equally important, discussions on the structure of AGN over the years.

This work is dedicated to Giorgio G. C. Palumbo, who taught MG the most important things.
\end{acknowledgements}

\bibliographystyle{aa}
\bibliography{myref}

\end{document}